\documentclass{iopart}
\usepackage{iopams}

\expandafter\let\csname equation*\endcsname\relax

\expandafter\let\csname endequation*\endcsname\relax

\usepackage{amsmath}
\usepackage{epsf,amssymb,fleqn}

\catcode`\@=11
\usepackage{pdfsync}
\usepackage{dsfont}

\usepackage[active]{srcltx}

\usepackage{epsfig}
\usepackage{epstopdf}
\usepackage{float}

\def\II{\hbox{{1}\kern-.25em\hbox{l}}}

\begin{document}
\begin{flushright}
{{{{{{{{\textsl{DESY 16--232}}}}}}}}}
\end{flushright}

\title[Gustafson integrals for $SL(2,\mathbb{C})$ spin magnet.]{  Gustafson integrals for $SL(2,\mathbb{C})$ spin magnet.}
\author{S. {\'E}. Derkachov~$^1$\,,
A. N. Manashov~$^{2,3}$ and P.~A. Valinevich~$^1$
}

\address{$^1$ St.Petersburg Department of Steklov
Mathematical Institute of Russian Academy of Sciences,\\
\ \ \
Fontanka 27, 191023 St.Petersburg, Russia.
 }

\address{$^2$  Institut f\"ur Theoretische Physik, Universit\"at Hamburg,   D-22761 Hamburg, Germany}
\address{$^3$ Institute for Theoretical Physics, University of  Regensburg, D-93040 Regensburg, Germany}

\ead{derkach@pdmi.ras.ru}
\ead{alexander.manashov@desy.de}
\ead{valinevich@pdmi.ras.ru}
\begin{abstract} It was observed recently that the multidimensional Mellin--Barnes integrals (Gustafson's integrals)
arise naturally in  studies of the $SL(2,R)$ spin chain models. We extend this analysis to the noncompact
$SL(2,\mathbb{C})$ spin magnets and obtain integrals which generalize Gustafson's integrals to the complex case.
\end{abstract}


\maketitle

\setcounter{footnote}{0}

\section{Introduction}

It was shown by E.~K.~Sklyanin~\cite{Sklyanin:1991ss} that the  eigenfunctions of the  monodromy matrix provide
convenient bases for studies of  spin chain magnets. In many cases these eigenfunctions can be constructed in
explicit form. Rather (un)expectedly the most simple and elegant expressions arise for the models with infinite
dimensional Hilbert spaces. Such models  include the so-called noncompact spin magnets and famous Toda chain. The
eigenfunctions constructed with the help of Quantum Inverse Scattering
Method~\cite{FST,TaFa,KulSk,Sklyanin:1995bm,Fad} (QISM)  are given by multi-parametric integrals which have a
hierarchical structure and can be represented as Feynman diagrams of certain type
\cite{Derkachov:2001yn,Derkachov:2002tf,Derkachov:2003qb,Silantyev07,Belitsky:2014rba}. In many cases the
calculation of scalar products between eigenfunctions or matrix elements can be, quite effectively, carried out on
the diagram level. For the Toda chain or the $SL(2,R)$ spin chains  the result is given, as a rule, by a product of
Euler's gamma functions depending on spectral parameters (separated variables). It was noticed
recently~\cite{Derkachov:2016dhc} that using the completeness condition for the eigenfunctions of the $SL(2,R)$ spin
magnets one can show that certain relations between scalar products and matrix elements take the form of
multidimensional Mellin--Barnes integrals which are equivalent to  the integrals derived by R.~A.~Gustafson
in~\cite{Gustafson,Gustafson92}.

In this work we apply the same program to the $SL(2,\mathbb{C})$ spin magnets in order to derive the counterparts of
the Gustafson's integrals in the complex case. The eigenfunctions of the monodromy matrix for the
$SL(2,\mathbb{C})$ magnet and the corresponding Sklyanin's measures were obtained in~\cite{Derkachov:2001yn,Derkachov:2014gya}.
Using these results and calculating matrix elements of the shift operator we derive an analog of the first
Gustafson's integral (Eq. (5.2) in Ref.~\cite{Gustafson}). It has exactly the same functional form. The only changes
amount  to a modification of the integration measure  and   the replacement of all
Euler gamma functions (the gamma-function associated with the real field
$\mathbb{R}$ in the classification of Ref.~\cite{GelfandGraevRetakh04})
entering this integral  by the
gamma functions associated with the complex field $\mathbb{C}$~\cite{GelfandGraevRetakh04}. It allows one to suggest
that all Gustafson's integrals admit the corresponding generalization.

The paper is organized in the following way. In sect.~\ref{sect:preliminaries} we recall the formulation of the
$SL(2,\mathbb{C})$ spin chain model and necessary facts from the QISM and SoV approach. In sect.~\ref{sect:integrals}
we calculate the relevant matrix elements and derive an analog of the first Gustafson's integral. Elements of the
diagrammatic technique are given in ~\ref{sect:Diagram}. Finally, we present the Mellin--Barnes form of the
star--triangle relation in \ref{sect:MST}.

\section{$SL(2,\mathbb{C})$ magnet}\label{sect:preliminaries}

The quantum $SL(2,\mathbb{C})$ spin magnet is a generalization of the ordinary $\mathrm{XXX}_{s}$ spin chain. The
dynamical variables of the model  are  spin generators which belong, at each site, to a unitary continuous principal
series representation of the $SL(2,\mathbb{C})$ group. Such a representation,
$T^{(s,\bar s)}$, is
determined by two complex  spins,
 $s$ and $\bar s$,  which are parameterized by a (half)integer number $n_s$ and a real number $\nu_s$~\cite{Gelfand}
\begin{align}
s=\frac{1+n_s}2+i\nu_s, && \bar s=\frac{1-n_s}2+i\nu_s.
\end{align}
The group transformation takes the form
\begin{align}\label{Tg}
[T^{(s,\bar s)}(g)\phi](z,\bar z)=(a-cz)^{-2s}(\bar a -\bar c \bar z)^{-2\bar s}\,\phi\left(\frac{dz-b}{a-cz},
\frac{\bar d \bar z-\bar b}{\bar a -\bar c \bar z}\right)\,,
\end{align}
where $g$ is a complex unimodular matrix,
$g=
\left(
\begin{array}{cc}
     a&b\\
c& d
\end{array}
\right)
$.
The transformation~(\ref{Tg}) is a unitary transformation on $L_2(\mathbb{C})$
\begin{align}\label{sc}
\langle \phi\, | \psi \rangle = \int d^2z\,
\overline{\phi(z,\bar z)}\,\psi(z,\bar z)\,,&&
\langle\, T^{(s,\bar s)}(g)\phi\, |\, T^{(s,\bar s)}(g)\psi\, \rangle = \langle \phi\, | \psi \rangle.
\end{align}
The generators of infinitesimal transformations (spin operators) take the form
\begin{align}\label{spin-operators}
S_- & =-\partial_z, && S_0 =z\partial_z+s,  && S_+=z^2\partial_z+2s\, z\,,
\nonumber\\
\bar S_- & =-\partial_{\bar z}, && \bar S_0 ={\bar z}\partial_{\bar z}+\bar s, && \bar S_+=\bar z^2\partial_{\bar z}+2\bar s\, \bar z\,.
\end{align}
They are adjoint to each other up to a sign, $S_\alpha^\dagger = -\bar S_\alpha$, and satisfy the $sl_2$ 
commutation relations
\begin{align}
[S_+,S_-]=2S_0 , && [S_0 ,S_\pm]=\pm S_\pm\,.
\end{align}
The anti-holomorphic generators satisfy exactly the same  relations. Henceforth, if  holomorphic and
anti-holomorphic equations are the same we will write down only the holomorphic version.

\subsection{$L$ operators and monodromy matrices}

The Hilbert space of the model is given by a direct product of  $N$ copies of the $L_2(\mathbb{C})$ space,
\begin{align}\label{HN}
\mathbb{H}_N=\mathbb{V}_1\otimes \mathbb{V}_2\otimes\cdots\otimes \mathbb{V}_N,  \qquad \mathbb{V}_k=L_2(\mathbb{C})\,,\qquad
k=1,\ldots,N.
\end{align}
We will consider only the homogeneous chains, i.e.  the spin generators~(\ref{spin-operators}) at each site have the
same spins,  $s_k=s$, $\bar s_k=\bar s$ for all $k$ and, for simplicity, we will assume that $ s-\bar s=n_s$ is an
integer number.

In the  QISM approach one defines (at each site) the so-called $\mathrm{L}$-operator
\begin{align}
\label{L}
L_k(u) =
\left (\begin{array}{cc}
u + i S^{(k)}_0 & i S_{-}^{(k)} \\
i S_{+}^{(k)} & u - i S^{(k)}_0 \end{array} \right )\,, &&
\bar L_k(\bar u) = \left(\begin{array}{cc}
\bar u + i \bar S_0^{(k)} & i \bar S_{-}^{(k)}\\
i \bar S_{+}^{(k)}& \bar u - i \bar S_0^{(k)}
\end{array}\right)
\end{align}
and constructs a monodromy matrix as a product of the $\mathrm{L}$-operators,
\begin{align}\label{monodromy}
T(u)=L_1(u)L_2(u)\ldots L_N(u)=\left(\begin{array}{cc}
A_N(u)& B_N(u)\\
C_N(u)& D_N(u)
\end{array}\right)
\,.
\end{align}
The anti-holomorphic monodromy matrix is given by the same expression with
$L_k(u)\to\bar L_k(\bar u)$. Note, that we do not assume any relation between  $u$ and $\bar u$, which are two independent parameters.

It is shown in the QISM~\cite{Fad,KulSk,Sklyanin:1991ss} that the entries of the monodromy matrix form commuting
operator families, i.e.
\begin{align}\label{AA}
[A_N(u), A_N(v)]=0, &&  [B_N(u), B_N(v)]=0, &&[C_N(u), C_N(v)]=0, &&  [D_N(u), D_N(v)]=0\,.
\end{align}
Moreover, in the case under consideration the operators $A_N$ and $D_N$ ($B_N$ and $C_N$) are related by the
inversion transformation~\cite{Derkachov:2014gya}. In what follows we consider the operators $A_N$ and $B_N$ only.

These operators commute with the total generators $S_\alpha=S_\alpha^{(1)}+\ldots +S_\alpha^{(N)}$ as follows
\begin{align}
[S_0\,, A_N(u)]=0\,, &&
[S_-\,, B_N(u)] = 0. 
\end{align}
Since the operators $A_N(u), \, \bar A_N(\bar u)$ ($B_N(u), \, \bar B_N(\bar u)$) commute for different values of
the spectral parameters they can be diagonalized simultaneously and their eigenfunctions do not depend on the
spectral parameters
$u,\bar u$. These eigenfunctions play a distinguished role in the QISM formalism and form the basis of the so-called
Sklyanin's representation of the Separated Variables~\cite{Sklyanin:1995bm}. For the $SL(2,\mathbb{C})$ magnet the
eigenfunctions of $A_N$ and $B_N$ operators were constructed in Refs.~\cite{Derkachov:2001yn,Derkachov:2014gya}.  We
present the explicit expressions for them in the next subsection.
\subsection{SoV representation}

Since by  construction the operators $A_N(u)$ and $B_N(u)$ are polynomials of degree $N$ and $N-1$ in $u$,
respectively, their eigenvalues are polynomials in $u$ as well. It turns out quite convenient to label the
eigenfunction by the roots of its eigenvalue. We accept the following notations for the eigenfunctions -- $\Psi_A
({\boldsymbol{x}}|\boldsymbol{z})$ and $\Psi_B (p,{\boldsymbol{x}}|\boldsymbol{z})$.
\begin{itemize}
\item
 The eigenfunction
$\Psi_A ({\boldsymbol{x}}|\boldsymbol{z})$ satisfies the equations
\begin{eqnarray}\label{AbAN}
A_N(u)\,\Psi_A ({\boldsymbol{x}}|\boldsymbol{z}) &= & (u-x_1)\cdots(u-x_N)\,\Psi_A ({\boldsymbol{x}}|\boldsymbol{z})\,,
\nonumber\\
\bar A_N(\bar u)\,\Psi_A({\boldsymbol{x}}|\boldsymbol{z}) &= & (\bar u-\bar x_1)\cdots(\bar u-\bar x_N)
\,\Psi_A ({\boldsymbol{x}}|\boldsymbol{z})\,.
\end{eqnarray}
Here we introduced the shorthand notations $\boldsymbol{z}=\{z_1,\ldots,z_N\}$  and
$\boldsymbol{x}=\{x_1,\ldots,x_N\}$\,. The anti-holomorphic variables  $\bar x_k$ are adjoint to $x_k$, $\bar x_k=x_k^\ast$
and parameterized as follows
\begin{align}\label{x-n-nu}
x_k=-\frac{in_k}{2}+\nu_k\,, && \bar x_k=\frac{in_k}{2}+\nu_k\,,
\end{align}
where $\nu_k$ is a real number and  $n_k$ is an integer number.

\item The eigenfunctions of the operator   $B_N$ are determined by the equations
\begin{eqnarray}
B_N(u)\,\Psi_B (p,{\boldsymbol{x}}|\boldsymbol{z}) &= &p(u-x_1)\cdots(u-x_{N-1})\,\Psi_B (p,{\boldsymbol{x}}|\boldsymbol{z})\,,
\nonumber\\
\bar B_N(\bar u)\,\Psi_B (p,{\boldsymbol{x}}|\boldsymbol{z})&= &\bar p(\bar u-\bar x_1)\cdots(\bar u-\bar x_{N-1})\,
\Psi_B (p,{\boldsymbol{x}}|\boldsymbol{z})\,,
\end{eqnarray}
where $\bar p =p^\ast$ and $\boldsymbol{x}=\{x_1,\ldots,x_{N-1}\}$ and the separated variables have the
form~(\ref{x-n-nu}).

\end{itemize}

The eigenfunctions of both operators can be constructed recursively~\cite{Derkachov:2001yn,Derkachov:2014gya}.
Namely, let us define two (layer) operators, $\Lambda_k(x)$ and $\widetilde \Lambda_k(x)$ which map  functions of
$k-1$ variables to  functions of $k$ variables.
The operator $\Lambda_k(x)$ is an integral operator defined as follows~\cite{Derkachov:2014gya}
\begin{align}
\big[\Lambda_k(x) \Phi\big](z_1,\ldots,z_k)&= r_k(x) \prod_{i=1}^{k-1}[z_i-z_{i+1}]^{1-2s}
\notag\\
&\quad \times\prod_{i=1}^{k-1}
\int d^2 w_i\,[w_i-z_i]^{s+ix-1}[w_i-z_{i+1}]^{s-ix-1}\,
\,\Phi(w_1\,,\ldots\,,w_{k-1})\,.
\end{align}
Here $[z]^{\alpha}\equiv z^\alpha \bar z^{\bar\alpha}$, the  parameter  $x$ has the form~(\ref{x-n-nu}) and the
normalization factor $r_k(x)$ is given by
\begin{align}
r_k(x)=\big(a(s+ix) a(\bar s-i\bar x)\big)^{k-1}\,, && a(\alpha)=\frac{\Gamma(1-\bar\alpha)}{\Gamma(\alpha)}.
\end{align}
The definition of the second  operator
  $\widetilde \Lambda_k(x)$
reads
\begin{equation}
\widetilde\Lambda_k(x) \equiv [z_k]^{-s+ix} \Lambda_k(x)=  \Lambda_k(x)[z_k]^{-s+ix}\,.
\end{equation}
The eigenfunctions can be written in the following explicit form~\cite{Derkachov:2014gya}
\begin{eqnarray}\label{reprPsiB}
\Psi_B (p\,,{\boldsymbol{x}}|\boldsymbol{z})  & = &|p|^{N-1}
\Lambda_{N}\left(x_1\right)\cdots
\Lambda_{2}(x_{N-1})\,\mathrm{e}^{i p z_1+i \bar{p} \bar{z}_1}\,,\\[2mm]
\label{reprPsiA}
\Psi_A ({\boldsymbol{x}}|\boldsymbol{z})  &=& \widetilde \Lambda_{N}\left(x_1\right)\cdots \widetilde \Lambda_{2}(x_{N-1})\,
\widetilde \Lambda_{1}(x_{N})\,.
\end{eqnarray}
The normalization of the layer operators is chosen in such a way that they satisfy the exchange relation,
$\Lambda_k(x_1)\Lambda_{k-1}(x_2)=\Lambda_k(x_2)\Lambda_{k-1}(x_1)$, (and similar for $\widetilde \Lambda$) that ensures that
the eigenfunctions~(\ref{reprPsiB}) are symmetric function of the separated variables $\boldsymbol{x}$, see
Ref.~\cite{Derkachov:2001yn,Derkachov:2014gya} for details.

\begin{figure}[t]
\centerline{\includegraphics[width=0.990\linewidth]{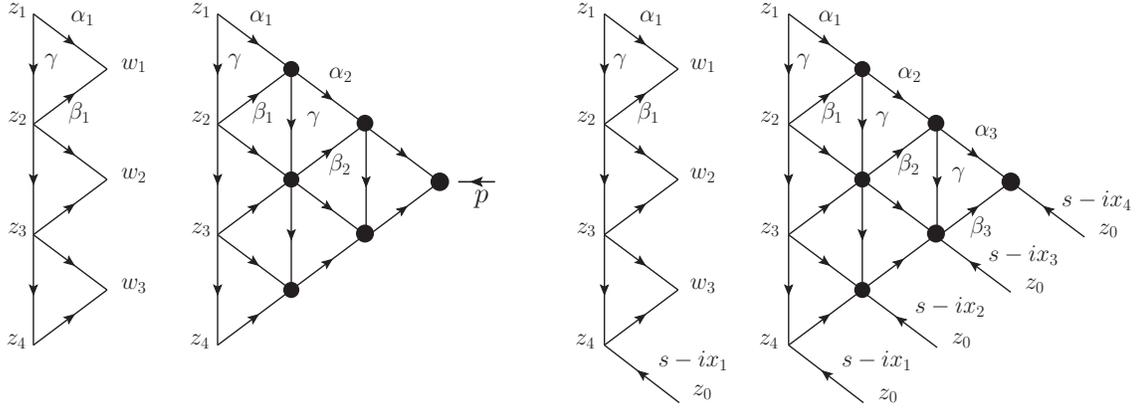}}
\caption{The diagrammatic representation for the layer operator $\Lambda_N(x_1)$ and
the eigenfunction
$\Psi_B(p,\boldsymbol{x}|\boldsymbol{z})$ for $N=4$ (two left figures)  and
the layer operator $\widetilde \Lambda_N(x_1)$ and the eigenfunction
$\Psi_A(\boldsymbol{x}|\boldsymbol{z})$ for $N=4$ (two right figures).
The line from $w$ to $z$ and the index $\alpha$ stands for the propagator, $[z-w]^{-\alpha}$, and the black dots -- for
the integration vertices.
The indices have the following values: $\alpha_k=1-s-i x_k$, $\beta_{k}=1-s+i x_k$
and $\gamma=2s-1$. For the operator $\widetilde \Lambda$ and the eigenfunction $\Psi_A(\boldsymbol{x}|\boldsymbol{z})$ the variable $z_0=0$. }
\label{PsiAB}
\end{figure}

It appears quite useful to represent the kernels of operators and  eigenfunctions as Feynman diagrams. Several  such diagrams
are shown in Fig.~\ref{PsiAB}. For more examples of the diagrammatic technique see Ref.~\cite{Derkachov:2001yn}.
Taking into account that
$\widetilde \Lambda_1(x_N)=[z_N]^{-s+ix_N}$ one can write the eigenfunction of the operator $A_N$ in the equivalent
form
\begin{eqnarray}\label{reprPsiA1}
\Psi_A ({\boldsymbol{x}}|\boldsymbol{z}) =
\Lambda_{N-1}\left(x_1\right)\cdots
\Lambda_{2}(x_{N-1})\,[z_N]^{-s+ix_1}\cdots [z_1]^{-s+ix_N}\,.
\end{eqnarray}
Using this representation and taking into account that the shift operator $T_{z_0}=e^{z_0 S_- + \bar z_0 \bar S_0}$
($T_{z_0} \Phi(\boldsymbol{z})=\Phi(z_1-z_0,\ldots,z_N-z_0)$) commutes with the layer operators one easily finds
that
\begin{eqnarray}\label{reprPsiAT}
T_{z_0}\Psi_{A}({\boldsymbol{x}}|\boldsymbol{z}) &=&
\Lambda_{N-1}\left(x_1\right)\cdots
\Lambda_{2}(x_{N-1})\,[z_{N}-z_0]^{-s+ix_1}
\cdots [z_{1}-z_0]^{-s+ix_N}
\nonumber\\
&=&\widetilde \Lambda_{N}^{(z_0)}\left(x_1\right)\cdots \widetilde \Lambda_{2}^{(z_0)}(x_{N-1})\,
\widetilde \Lambda_{1}^{(z_0)}(x_{N})\,,
\end{eqnarray}
where $\Lambda_{k}^{(z_0)}=\Lambda_k(x)[z_k-z_0]^{-s+ix}$. The diagram for  the function~(\ref{reprPsiAT}) is shown
in Fig.~\ref{PsiAB}. 

\vskip 2mm

 The functions $\Psi_A({\boldsymbol{x}}|\boldsymbol{z})$ and $\Psi_B(p\,,{\boldsymbol{x}}|\boldsymbol{z})$
being eigenfunctions of the self-adjoint operators form a complete orthogonal basis in the Hilbert space of the
model
\begin{eqnarray}\label{SCS}
\int {d}^{2N}\boldsymbol{z}\,
\overline{\Psi_A({\boldsymbol{x}'}|\boldsymbol{z})}\, \Psi_A({\boldsymbol{x}}|\boldsymbol{z}) &=&
                \big(\boldsymbol{ \mu}^{(A)}_N(\boldsymbol{x})\big)^{-1}\,\delta_N(\boldsymbol{x}-\boldsymbol{x}')\,,
                \\
\int {d}^{2N}\boldsymbol{z}\,
\overline{\Psi_B(p'\,,{\boldsymbol{x}'}|\boldsymbol{z})}\,  \Psi_{B} (p\,,{\boldsymbol{x}}|\boldsymbol{z}) &=&
                         \big(\boldsymbol{\mu}^{(B)}_N(\boldsymbol{x})\big)^{-1}\,
                           \delta^2(\vec{p}-\vec{p}')\,
\delta_{N-1}(\boldsymbol{x}-\boldsymbol{x}')\,.
\end{eqnarray}
Here $d^{2N}\!\boldsymbol{z}=d^2z_1\ldots d^2z_N$ and
the delta function $\delta_N(\boldsymbol{x}-\boldsymbol{x}')$ is defined as follows:
\begin{eqnarray}
\delta_N(\boldsymbol{x}-\boldsymbol{x}')=\frac1{N!}
\sum_{s\in S_N}\delta^{(2)}(x_1-x'_{s(1)})\cdots
\delta^{(2)}(x_N-x'_{s(N)})\,,
\end{eqnarray}
where the sum goes over all permutations of $N$ elements and
\begin{equation}
\delta^{(2)}(x-x')\equiv
\delta_{n n'} \delta(\nu-\nu').
\end{equation}
The weight functions
$\boldsymbol{ \mu}_N(\boldsymbol{x})$ and $\boldsymbol{ \mu}_N(p,\boldsymbol{x})$ are the so-called
Sklyanin's measures. They were calculated in Refs.~\cite{Derkachov:2001yn,Derkachov:2014gya} and take the following
form
\begin{eqnarray}
\boldsymbol{\mu}^{(A)}_N(\boldsymbol{x}) &= &
\frac1{N!}\frac{\pi^{-N^2}}{(2\pi)^{N}}  \prod_{k<j}[x_k-x_j]\,,
\\
\boldsymbol{\mu}^{(B)}_N(\boldsymbol{x}) &=&\frac1{(N-1)!} \frac{2\pi^{-N^2}}{(2\pi)^{N}}
\prod_{k<j}
[x_k-x_j]\,.
\end{eqnarray}
The completeness condition for the functions
$\Psi_A({\boldsymbol{x}}|\boldsymbol{z})$ and
$\Psi_B({\boldsymbol{x}}\,,{\boldsymbol{p}}|\boldsymbol{z})$ reads
\begin{eqnarray}\label{completA}
 \prod_{k=1}^N\delta^{(2)}(\vec{z}_k-\vec{z}'_k) &=&\int \mathcal{D}_N \boldsymbol{x}\,\boldsymbol{\mu}^{(A)}_N(\boldsymbol{x})
\Psi_A({\boldsymbol{x}}|\boldsymbol{z})\,
\overline{\Psi_A({\boldsymbol{x}}|\boldsymbol{z}')}\,,\\
\label{completB}
\prod_{k=1}^{N}\delta^{(2)}(\vec{z}_k-\vec{z}'_k)&=&
\int d^2 p \,\int \mathcal{D}_{N-1} \boldsymbol{x}\,\boldsymbol{\mu}^{(B)}_N(\boldsymbol{x})
\Psi_B({\boldsymbol{x}}\,,{\boldsymbol{p}}|\boldsymbol{z})\,
\overline{\Psi_B({\boldsymbol{x}}\,,{\boldsymbol{p}}|\boldsymbol{z}')} \,,
\end{eqnarray}
where the symbol $\mathcal{D}_N\boldsymbol{x}$ stands for
\begin{eqnarray}
\int \mathcal{D}_N\boldsymbol{x}=\prod_{k=1}^N \left(\sum_{n_k=-\infty}^{\infty}\int_{-\infty}^{\infty} d\nu_k\right)\,
\end{eqnarray}
and the sum  goes over all integers. The relations (\ref{completA}), (\ref{completB}) can be easily checked for
$N=1,2$. The proof for general $N$  will be given elsewhere.

Let us note that for the case $N=1$ the eigenfunctions of the operators
$B_1 = -i\partial_z$  and $A_1(u)= u + i\big(s+z\partial_z\big) $
are the exponential and power functions, respectively. Namely,
\begin{align}
\Psi_B (p|z) = e^{ipz+i\bar p\bar z}, && \Psi_A (x|z) = [z]^{ix-s}\,.
\end{align}
The orthogonality and completeness relations for the power functions read
\begin{eqnarray}\label{N=1-completeness}
\int d^2z\, [z]^{ix_1-s} ([z]^{ix_2-s})^\ast &= &\int d^2z\, [z]^{-1+i(x_1-x_2)}=
 2 \pi^2 \delta^{(2)}(x_1-x_2)\,,\notag\\
\int Dx [z_1]^{ix-s} ([z_2]^{ix-s})^\ast & = &{[z_1]^{-s} ([z_2]^{-s})^*}\int Dx [z_1/z_2]^{ix}=2\pi^2\delta^2(\vec{z}_1-\vec{z}_2)\,
\end{eqnarray}
and agree with~(\ref{SCS}),~(\ref{completA}).

\section{Matrix elements and integrals identities }\label{sect:integrals}

\begin{figure}[t]
\centerline{\includegraphics[width=0.890\linewidth]{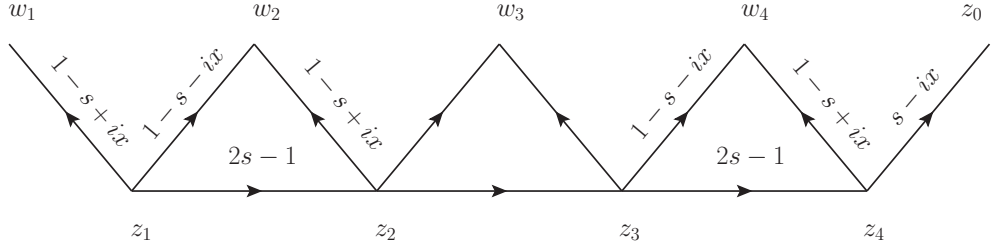}}
\caption{The diagrammatic representation for the operator $Q_N^{(z_0)}(x)$ for $N=4$.
Note that the index of the rightmost arrow differs from others.}
\label{fig:QX}
\end{figure}

Let us calculate  the matrix element of the shift operator
 between the eigenstates of the  operator $A_N$. We define
\begin{equation}\label{Txxdef}
T_{z_0}(\boldsymbol{x},\boldsymbol{x}^\prime)=\langle \Psi_A(\boldsymbol{x}')|T_{z_0}|\Psi_A(\boldsymbol{x})\rangle\,.
\end{equation}
The calculation of (\ref{Txxdef}) goes along the following lines: first, using the representations~(\ref{reprPsiA})
and (\ref{reprPsiAT}) we write the matrix element as follows
\begin{equation}
\langle \Psi_A(\boldsymbol{x}')|T_{z_0}|\Psi_A(\boldsymbol{x})\rangle=
\widetilde \Lambda^\dagger_{1}(x'_{N})  \widetilde \Lambda_{2}^\dagger(x'_{N-1}) \cdots\widetilde \Lambda_{N}^\dagger(x'_1)\,
\widetilde \Lambda_{N}^{(z_0)}\left(x_1\right)\cdots \widetilde \Lambda_{2}^{(z_0)}(x_{N-1})\,
\widetilde \Lambda_{1}^{(z_0)}(x_{N})\,.
\end{equation}
Second, representing the product $\widetilde \Lambda_{N}^\dagger(x'_1)\, \widetilde \Lambda_{N}^{(z_0)}(x_1)$ as a
Feynman diagram and simplifying it with the help of the identities (\ref{Chain}), (\ref{Star}), (\ref{Cross})  in
\ref{sect:Diagram} one gets
\begin{equation}
\widetilde \Lambda_{N}^\dagger(x')\, \widetilde \Lambda_{N}^{(z_0)}(x)
=[z_0]^{i(x-x')} (-1)^{[s-ix]  } 
\,q(x,x')\, (Q^{(z_0=0)}_{N-1}(x))^\dagger Q^{(z_0)}_{N-1}(x')\,,
\end{equation}
 where the sign factor $(-1)^{[a]}\equiv (-1)^{a-\bar a}$ and the diagram for the operator
$Q^{(z_0)}_{N}(x)$ is shown in Fig.~\ref{fig:QX}.   The factor $q(x,x')$ is given by the following expression
\begin{equation}\label{q-f-def}
q(x,x')=\pi a\big(1+i(x-x')\big) \frac{a(\bar s- i\bar x)}{a(s-ix')}\,.
\end{equation}
Third, using the diagrammatic technique it is straightforward to check that
\begin{equation}
Q^{(z_0)}_{k}(x') \widetilde \Lambda_{k}^{(z_0)}(x)=q(x,x') \widetilde \Lambda_{k}^{(z_0)}(x)Q^{(z_0)}_{k-1}(x')\,.
\end{equation}
Thus one  can reduce the $N$-point scalar product to the $N-1$-point product multiplied by some factor. It allows
one  immediately to get an answer for the matrix element~(\ref{Txxdef})
\begin{equation}\label{Txx-result}
T_{z_0}(\boldsymbol{x},\boldsymbol{x}^\prime)=(-1)^{[A_X]}[z_0]^{i(X-X')}\prod_{k,j=1}^{N}
q(x_k,x'_j)\,,
\end{equation}
where we introduced the notations
\begin{align*}
X=\sum_{k} x_k, &&\bar X=\sum_k \bar x_k, && A_X 
=\sum_k (s-ix_k), &&
\bar A_{\bar X}
=\sum_k (\bar s-i\bar x_k).
\end{align*}

The calculation of the scalar product between the eigenfunctions of the operators $A_N$ and $B_N$  follows exactly
the same lines so we give the final answer only
\begin{equation}\label{BA-result}
\langle \Psi_B(p,\boldsymbol{u})|\Psi_A(\boldsymbol{x})\rangle =
       i^{[A_X]} {\pi^N}{|p|^{-N-1}} [p]^{A_X}
     \prod_{k=1}^{N}a(\bar s-i\bar x_k) \prod_{j=1}^{N-1}
q(x_k,u_j)\,.
\end{equation}
Note that the expressions (\ref{Txx-result}) and  (\ref{BA-result})  are symmetric functions of the separated
variables as it should be. We also remark here that Eqs.~(\ref{Txx-result}),  (\ref{BA-result}) have  striking
resemblance to the analogous expressions in the $SL(2,R)$ spin chain models, see
Refs.~\cite{Belitsky:2014rba,Derkachov:2016dhc}.

The function $a(1+i(x-x'))$ entering (\ref{q-f-def}) becomes singular for $x=x'$. Indeed,
\begin{align}
a(1+i(x-x')) &=\frac{\Gamma(i(\bar x'-\bar x))}{\Gamma(1+i(x-x'))}
\notag\\
& =\frac{\Gamma(i(\nu'-\nu)+(n-n')/2)}{\Gamma(1-i(\nu'-\nu)+(n-n')/2)}
 =(-1)^{n-n'}\frac{\Gamma(i(\nu'-\nu)-(n-n')/2)}{\Gamma(1-i(\nu'-\nu)-(n-n')/2)}\,.
\end{align}
Thus the function  $a(1+i(x-x'))$ is singular only when $n=n'$ and $\nu=\nu'$. The divergency comes from the chain
integration and the right way to regularize it is to give the variable $\nu$ a small imaginary part, i.e.
$x=-in/2+ i\nu,\, \bar x = in/2+i\nu$ where $\text{Im}\,\nu>0$ (Note, that the variable $x$ is related to the function
on the right side of the scalar product). So from now on we assume, whenever it is necessary, that the parameters
$\nu_k$ in   Eq.~(\ref{x-n-nu})  have a positive imaginary part.

\subsection{Gustafson integrals for $SL(2,\mathbb{C})$}
In this section we present a generalization of the Gustafson's integrals (Eq.~5.2 in Ref.~\cite{Gustafson}) to the
complex case. Using the completeness condition~(\ref{completB}) for the $B$--system one can represent the matrix
element~(\ref{Txxdef}) in the form
\begin{equation}\label{TBA}
T_{z_0}(\boldsymbol{x},\boldsymbol{x}')=\int d^2 p \, e^{-ipz_0-i\bar p\bar z_0}
\int \mathcal{D}_{N-1} \boldsymbol{u}\,\boldsymbol{\mu}^{(B)}_N(\boldsymbol{u})
\overline{\langle \Psi_B(p,\boldsymbol{u})|\Psi_A(\boldsymbol{x}')\rangle}
\langle \Psi_B(p,\boldsymbol{u})|\Psi_A(\boldsymbol{x})\rangle\,,
\end{equation}
where we take into account that $T_{z_0} \Psi_B(p,{\boldsymbol{u}})= e^{-ipz_0-i\bar p\bar
z_0}\Psi_B(p,{\boldsymbol{u}})$.
 Substituting the expressions~(\ref{Txx-result}) and  (\ref{BA-result}) into (\ref{TBA}) one gets after some algebra
\begin{multline}\label{C-gustafson}
\frac1{(N-1)!}\left(\prod_{k=1}^{N-1}\sum_{m_k=-\infty}^{\infty}\int_{-\infty}^{\infty} \frac{d\nu_k}{2\pi}\right)
\frac{\prod_{k=1}^N \prod_{j=1}^{N-1} a(1+i(x_k-u_j))\,a(1+i(u_j-x'_k))}{\prod_{m<j} a(1+i(u_j-u_m))a(1+i(u_m-u_j))} =
\\
=\frac{\prod_{k,j=1}^N a(1+i(x_k-x'_j))}{a(1+i(X-X'))}\,.
\end{multline}
We recall here that the integration variables $u_k$ take the values: $u_k=-{in_k}/2+\nu_k,\ \ \bar u_k=
{in_k}/2+\nu_k $, where $n_k$ is an integer and  $\nu_k$ is a real number. The external parameters $x_k,x'_k$ are
\begin{align*}
x_k=-\frac{im_k}2+\mu_k, && \bar x_k= \frac{im_k}2+\mu_k, && x'_k=-\frac{im'_k}2+\mu'_k, && \bar x_k= \frac{im'_k}2+\mu'_k\,,
\end{align*}
where $m_k,m'_k$ are integers and $\mu_k$ and $\mu'_k$ are complex numbers such that $\text{Im}\,\mu_k>0$ and
$\text{Im}\,\mu'_k<0$. It can be checked that for such a prescription the $\nu$-poles of the
functions $ a(1+i(x_k-u_j)) $ and $a(1+i(u_j-x'_k))$ are separated by the integration contour.

We also recall that   the function $a(\alpha)\equiv a(\alpha,\bar\alpha))$, see Eq.~(\ref{a-def}),
 is   a function of two complex variables
such that $\alpha-\bar\alpha=n$. Namely,
$a(\alpha)={\Gamma(1-\bar\alpha)}/{\Gamma(\alpha)} $
and it is related to the gamma function for the complex field $\mathbb{C}$   defined in~\cite{GelfandGraevRetakh04}
\begin{align}\label{C-Gamma}
\Gamma(\alpha,\bar\alpha)=i^{\alpha-\bar\alpha} \frac{\Gamma(\alpha)}{\Gamma(1-\bar\alpha)}=i^{\alpha-\bar\alpha} a(1-\bar \alpha)\,.
\end{align}
Thus, Eq.~(\ref{C-gustafson}) is a direct analog of the first Gustafson's integral~(Eq.~(5.2) in
Ref.~\cite{Gustafson}) --- the only difference consists in replacing the Euler gamma function by the function
(\ref{C-Gamma}) and the corresponding modification of the integration measure. In was shown
in~\cite{Derkachov:2016dhc} that many of Gustafson's integrals can be obtained from the analysis of the matrix
elements of the $SL(2,R)$ spin chain models. There is little doubt that such an analysis can be extended to the
$SL(2,\mathbb{C})$ magnet and, therefore, it seem very plausible that many of  Gustafson's integrals admit an
extension to the complex case.


\ack This study was supported by the Russian Science Foundation (S.~D. and P.~V.), project
$\text{N}^{\text{o}}$ 14-11-00598,  and by Deutsche Forschungsgemeinschaft (A.~M.),  grant MO~1801/1-1.

\appendix

\addcontentsline{toc}{section}{Appendices}

\renewcommand{\theequation}{\Alph{section}.\arabic{equation}}
\renewcommand{\thetable}{\Alph{table}}
\setcounter{section}{0} \setcounter{table}{0}
\section{Diagram technique}\label{sect:Diagram}
This Appendix contains   elements of the diagram technique which were used throughout the paper. The functions and
kernels of operators   are represented in the form of two-dimensional Feynman diagrams. The propagator is shown by
the arrow directed from $w$ to $z$ with the index
$\alpha$ attached to~it

\begin{figure}[H]
\centerline{\includegraphics[width=0.35\linewidth]{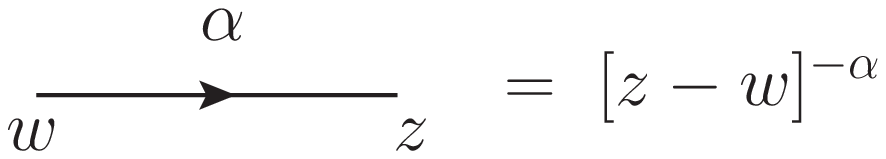}}
\end{figure}
\noindent The propagator is given by the following expression
\begin{equation}
\frac{1}{[z-w]^\alpha}\equiv\frac{1}{(z-w)^\alpha (\bar z-\bar w)^{\bar\alpha}}=
\frac{(\bar z-\bar w)^{\alpha-\bar\alpha}}{|z-w|^{2\alpha}}=\frac{(-1)^{\alpha-\bar\alpha}}{[w-z]^{\alpha}}\,,
\end{equation}
where $\alpha-\bar\alpha=n_\alpha$ is integer. Performing  the Fourier transform one defines the propagator in the
momentum representation
\begin{equation}\label{Fourier}
\int d^2 z e^{i(pz+\bar p\bar z)}[z]^{-\alpha}=\pi\, i^{\alpha-\bar\alpha}\, a(\alpha)\,{[p]^{\alpha-1}}\,.
\end{equation}
Here the notation $a(\alpha)$ is introduced for the function
\begin{equation}\label{a-def}
a(\alpha)\equiv a(\alpha,\bar\alpha)=\frac{\Gamma(1-\bar\alpha)}{\Gamma(\alpha)}\,, \quad a(\bar\alpha)=\frac{\Gamma(1-\alpha)}{\Gamma(\bar\alpha)}\,,
\quad a(\alpha,\beta,\gamma,\ldots)=a(\alpha)a(\beta) a(\gamma)\ldots
\end{equation}
It has the following properties
\begin{equation*}
a(\alpha) a(1-\bar\alpha)=1\,, \quad
a(1+\alpha) = -\frac{a(\alpha)}{\alpha\bar\alpha}\,, \quad
a(\alpha)a(1-\alpha)=(-1)^{\alpha-\bar\alpha}\,,\quad  a(\alpha)=(-1)^{\alpha-\bar\alpha} a(\bar\alpha)\,.
\end{equation*}
The evaluation of  Feynman diagrams is based on their transformation with the help of the certain rules
\begin{itemize}
\item Chain  relation:
\begin{equation}\label{Chain}
\int d^2 w\frac{1}{[z_1-w]^\alpha [w-z_2]^{\beta}}=
(-1)^{\gamma-\bar\gamma}a(\alpha,\beta,\gamma)
\frac{1}{[z_1-z_2]^{\alpha+\beta-1}}\,,
\end{equation}
where $\gamma=2-\alpha-\beta,\ \bar\gamma=2-\bar\alpha-\bar\beta$.
\begin{figure}[H]
\centerline{\includegraphics[width=0.7\linewidth]{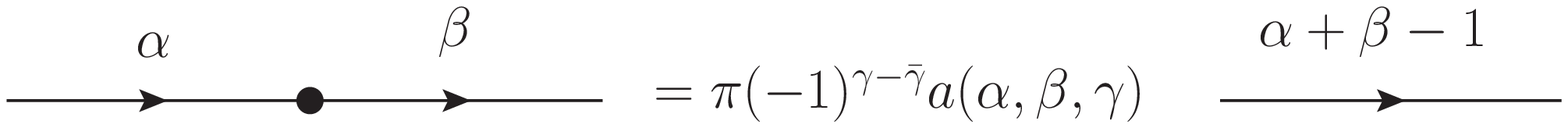}}
\end{figure}

\item Star--\,triangle relation:
\begin{equation}\label{Star}
\int d^2w\frac{1}{[z_1-w]^\alpha[z_2-w]^\beta [z_3-w]^\gamma}=
\frac{\pi a(\alpha,\beta,\gamma)}{[z_2-z_1]^{1-\gamma}[z_1-z_3]^{1-\beta}[z_3-z_2]^{1-\alpha}}\,,
\end{equation}
where $\alpha+\beta+\gamma=2$ and $\bar\alpha+\bar\beta+\bar\gamma=2$.
\begin{figure}[H]
\centerline{\includegraphics[width=0.7\linewidth]{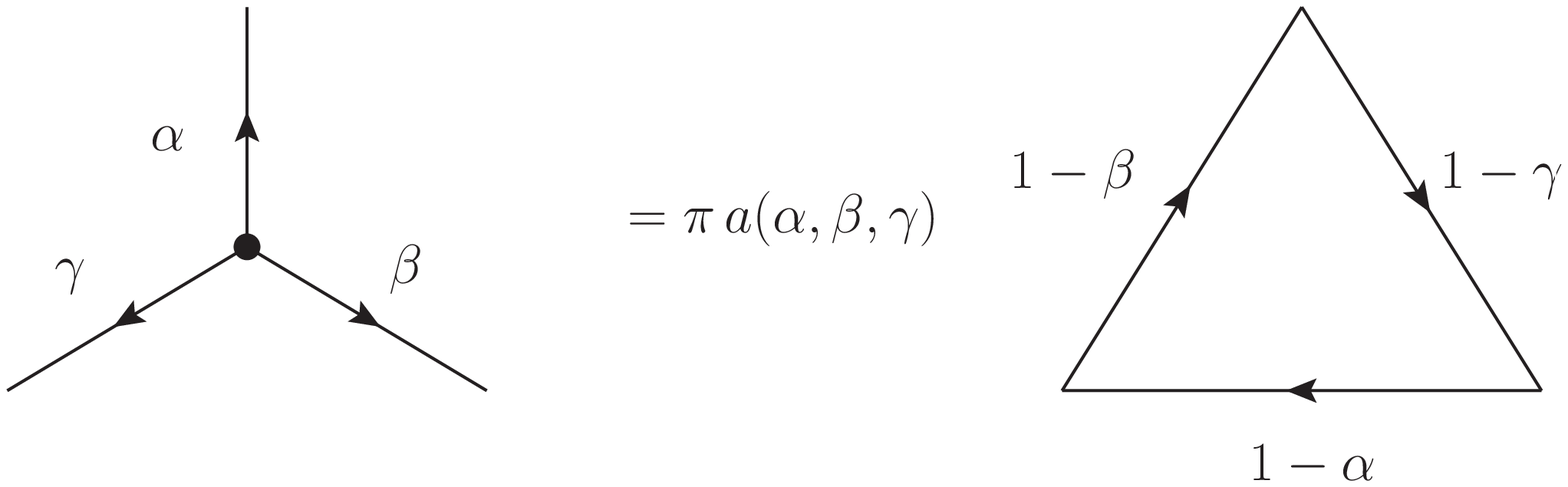}}
\end{figure}
\item Cross relation:
\begin{eqnarray}\label{Cross}
\frac{1}{[z_1-z_2]^{\alpha'-\alpha}}\int d^2w
\frac{a(\alpha',\bar\beta')}{[w-z_1]^\alpha[w-z_2]^{1-\alpha'}
[w-z_3]^\beta [w-z_4]^{1-\beta'}}=
\nonumber\\
=\frac{1}{[z_3-z_4]^{\beta'-\beta}}
\int d^2\zeta
\frac{a(\alpha,\bar\beta)}{[w-z_1]^{\alpha'}[w-z_2]^{1-\alpha}
[w-z_3]^{\beta'} [w-z_4]^{1-\beta}}\,,
\end{eqnarray}
where $ s- \bar s\in \mathbb{Z}$ and  $\alpha+\beta=\alpha'+\beta'$.

\begin{figure}[H]
\centerline{\includegraphics[width=0.8\linewidth]{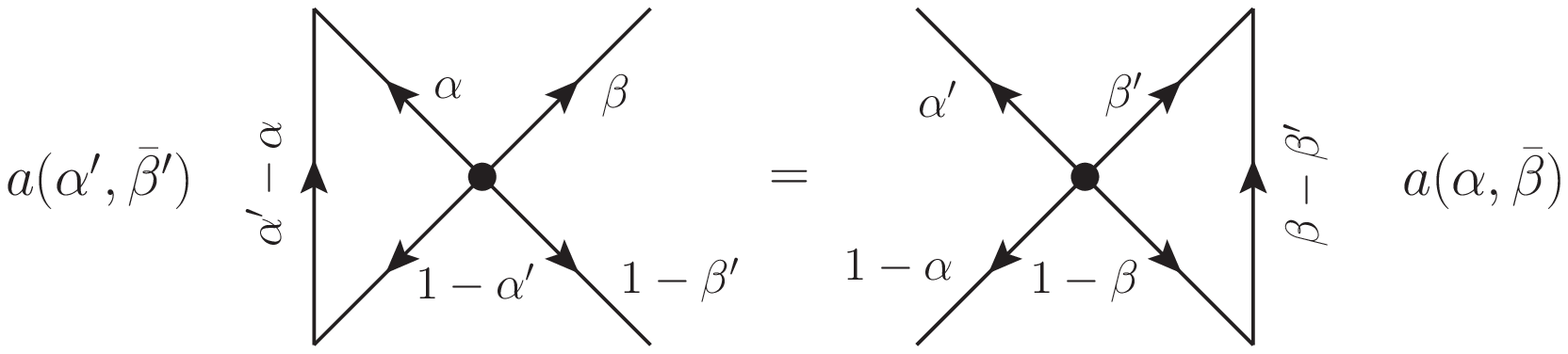}}
\end{figure}
\end{itemize}

\section{Mellin transform and star-triangle relation} \label{sect:MST}

In the simplest example $N=1$ (one site spin chain) the SoV-transformation related to the eigenfunctions of the
$A$-operator coincides with the Mellin--Barnes transformation. In this Appendix
we  show that the star-triangle identity~\cite{Bazhanov:2007vg} is equivalent to the star-triangle
relation~(\ref{Star}) in the Mellin--Barnes representation.

Let $f(z,\bar z)=f(r,\varphi)$  be a function on the complex plane. Combining the Fourier transform with respect to
the angle variable
$\varphi$ and the Mellin transform with respect to the radial variable~$r$ one gets
\begin{eqnarray}
f(z,\bar z)= \sum_{n=-\infty}^\infty e^{i\varphi n} \int_{-\infty}^{\infty} d\nu\, r^{-2i\nu-1} f_n(\nu) =
 \int D \alpha\,[z]^{-1/2-\alpha} \hat{f}(\alpha),
\end{eqnarray}
where $\alpha=i\nu-n/2$, $\bar \alpha =i\nu+n/2\,\,$  and
\begin{eqnarray}
\hat{f}(\alpha) \equiv f_n(\nu) = \frac{1}{2\pi^2} \int_0^{2\pi} d\varphi\,
e^{-i\varphi n} \int_0^{\infty} dr\, r^{2i\nu-1} f(r,\varphi)=
\frac1{2\pi^2}\int d^2 z [z]^{-1/2+\alpha}\,f(z,\bar z)\,.
\end{eqnarray}
These formulae are equivalent to the relations
\begin{align}\label{1}
\int d^2z\,[z]^{-1+\alpha} = 2 \pi^2 \delta^{(2)}(\alpha) \,,&&
\int D \alpha\,[z_1]^{-\alpha}[z_2]^{\alpha} =
2\pi^2[z_1]\,\delta^2(\vec{z}_1-\vec{z}_2)\,,
\end{align}
which are nothing else as the orthogonality and completeness relations~(\ref{N=1-completeness}). In order to avoid
misunderstanding we recall that
\begin{align*}
\int D \alpha = \sum_{n\in \mathbb{Z}}\int_{-\infty}^{+\infty}\mathrm{d}\nu, &&
\delta^{(2)}(\alpha-\alpha')\equiv
\delta_{n n'} \delta(\nu-\nu').
\end{align*}
Let us transform the start-triangle relation ($\beta_1+\beta_2+\beta_3=1$)
\begin{equation}\label{str}
\int d^2w \frac{1}{[z_1-w]^{1-\beta_1}[z_2-w]^{1-\beta_2} [z_3-w]^{1-\beta_3}}=
\frac{\pi a(1-\beta_1\,,1-\beta_2\,,1-\beta_3)}
{[z_2-z_1]^{\beta_3}[z_1-z_3]^{\beta_2}[z_3-z_2]^{\beta_1}}\,
\end{equation}
to the  Mellin--Barnes form. First of all, making use of the chain relation~(\ref{Chain}) we derive the following
representation for the propagator
\begin{eqnarray}
[w-z]^{\beta-1}=
\frac{1}{2\pi} a(1-\beta)\int D\alpha
\,\frac{a(1/2+\beta/2-\alpha) }{ a(1/2-\beta/2-\alpha) } \,[z]^{-1/2+\beta/2-\alpha}\,[w]^{-1/2+\beta/2+\alpha}\,,
\end{eqnarray}
Next, multiplying  both sides of Eq.~(\ref{str}) by the product
$\prod_{i=1}^3 [z_i]^{-1/2-\beta_i/2+\alpha_i}$ and integrating
over all variables $z_i$ with the help of Eq.~(\ref{1}) we obtain
\begin{multline}
(2\pi^2)\,\pi^3\,\delta^{(2)}\left(\sum \alpha_i\right)
\prod_{i=1}^3 a(1-\beta_i)\frac{a\left(1/2+{\beta_i}/2-\alpha_i\right)}
{a\left(1/2-{\beta_i}/2-\alpha_i\right)} =
\\
=\pi^4\,\prod_{i=1}^3 \int D \gamma_i \frac{a(1-\beta_i/2-\gamma_i)}{a(\beta_i/2-\gamma_i)}
\delta^{(2)}(\gamma_3-\gamma_2-\alpha_1)
\delta^{(2)}(\gamma_1-\gamma_3-\alpha_2)
\delta^{(2)}(\gamma_2-\gamma_1-\alpha_3)\,.
\end{multline}
Comparing the coefficient at the delta function $\delta^{(2)}\left(\sum \alpha_i\right)$ on both sides we get
%
\begin{eqnarray}
2\pi
\prod_{i=1}^3 a(1-\beta_i)\frac{a\left(1/2+{\beta_i}/2-\lambda_{i,i+1}\right)}
{a\left(1/2-{\beta_i}/2-\lambda_{i,i+1}\right)}
=\int D\gamma \prod_{i=1}^3\frac{a(1-\beta_i/2-\gamma+\lambda_{i-1})}
{a(\beta_i/2-\gamma+\lambda_{i-1})}\,,
\end{eqnarray}
where we put $\alpha_i=\lambda_{i}-\lambda_{i+1}$,
$\gamma_i=-\lambda_{i-1}+\gamma$, $\lambda_{i+3}\equiv\lambda_{i}$.
 For the special choice of the parameters this relation is reduced to the to the star-triangle identity, Eq.~(22)
in Ref.~\cite{Bazhanov:2007vg}.

\end{document}